# Ethical Design of Computers: From Semiconductors to IoT and Artificial Intelligence

**Sudeep Pasricha**, Colorado State University; **Marilyn Wolf**, University of Nebraska-Lincoln

**Abstract**: Computing systems are tightly integrated today into our professional, social, and private lives. An important consequence of this growing ubiquity of computing is that it can have significant ethical implications of which computing professionals should take account. In most real-world scenarios, it is not immediately obvious how particular technical choices during the design and use of computing systems could be viewed from an ethical perspective. This article provides a perspective on the ethical challenges within semiconductor chip design, IoT applications, and the increasing use of artificial intelligence in the design processes, tools, and hardware-software stacks of these systems.

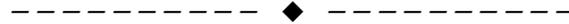

## 1 INTRODUCTION

The ethical design of computer systems is an important topic for hardware and software designers. Ethics refers to the study of right and wrong and is one of the oldest concerns of human thought. It prescribes what humans ought to do, usually in terms of rights, obligations, benefits to society, and fairness. Ethical principles guide us to strive for the virtues of honesty, compassion, and loyalty, while preserving crucial human rights, such as the right to life, the right to freedom from injury, and the right to privacy. These fundamental ethical principles, however, may need articulation to understand their application to computers.

How can we effectively integrate ethics into computer design activities? While computers offer great promise for better lives, they can also cause problems: design bugs can endanger human lives [1], computers can be used in crimes, and data security and privacy can be compromised by design flaws. However, eliminating computers or lobotomizing computer systems to eliminate their ability to perform certain functions is impractical. Computers perform many necessary functions in all aspects of society: finance, medicine, transportation, networking, and critical infrastructure. We need to find ways to understand and minimize their harmful consequences while maximizing the ability of computers to build better lives for people.

We can identify many categories of ethical challenges:
- Direct harm to people, animals, or property. Harm could be physical or cognitive/emotional.
- Systemic damage ranging from environmental pollution to encouragement of bad behavior.
- Waste and excessive use.
- A wide range of attack scenarios. The goals of an attacker may include fraud, direct or indirect damage, violations of security or privacy.

The Codes of Ethics of IEEE [2] and ACM [3] provide goals and expectations for engineering and computing professionals to begin to address these ethical challenges. Unfortunately, these codes are designed to frame expectations and do not provide direct guidance on how to act ethically in a specific situation during computer design. Thus, it is vital to understand the spectrum of ethical challenges that may arise during computer design and then explore actionable guidance to meet ethical design goals.

A distinction should also be made between the fields of engineering ethics and computing ethics, which are related but not entirely the same. Both relate to the creation of artifacts. However, the very broad set of applications enabled by software mean that computers can create a wide range of new problematic situations---bias in artificial intelligence (AI) based classification systems is one example. Moreover, engineering practices designed for the rapid deployment of consumer goods and services may not always be appropriate for safety-critical computing systems.

This article provides a perspective on the complex ethical challenges associated with computer design. In the next section, we outline the most important ethical challenges during computer design, including semiconductor fabrication, design of Internet of Things (IoT) devices, and the use of artificial intelligence (AI) in hardware/software stacks. Section 3 discusses approaches that are needed to address the outlined ethical challenges. Section 4 describes how ethics-centric computer design can be accomplished. Lastly, Section 5 presents some concluding thoughts.

## 2 ETHICAL COMPUTING CHALLENGES

In this section, we summarize some of the key ethical challenges facing computer system design. These challenges span the spectrum of the semiconductor industry including manufacturing and design of electronic chips, developers of mobile and IoT devices, and AI developers.

## 2.1 Semiconductor Fabrication

The design of electronic computing chips in semiconductor fabrication facilities involves many ethical concerns.

**Conflict minerals:** The mining of minerals such as cassiterite, columbite, tantalite, wolframite and their derivative elements tin, tantalum, tungsten, and gold, in some parts of the world, such as the Democratic Republic of the Congo (DRC) and adjoining countries, has supported conflict, human rights violations, and labor and environmental abuses [4]. Many of these minerals are essential for electronic chip design today. Semiconductor companies that source such minerals from these vulnerable parts of the world contribute directly or indirectly to abuses in these places. Thus, the sourcing and use of such minerals in electronic chip design is an important ethical concern.

**Semiconductor fabrication toxins:** Making computer chips involves the use of hundreds of chemicals. In the mid-1980s in the USA, women on the semiconductor production lines worked in cleanrooms and wore protective suits, but were still exposed to, and in some cases directly touched, chemicals that included reproductive toxins, mutagens, and carcinogens [5]. Later studies showed that women at various semiconductor plants had miscarriages at twice the expected rate due to exposure to toxins. Pledges were made to phase out the use of such chemicals in chipmaking to address this ethical failure. However, as chip production shifted to less expensive countries, studies showed that thousands of women and their unborn children In those countries continued to face exposure to the same toxins [5]. The semiconductor industry remains secretive about the use of such toxins across fabs in the global chipmaking ecosystem even to this day.

**Forced labor:** The global supply chains of electronics manufacturers have received much attention in recent years for the widespread problem of forced labor for the manufacture of components [6]. According to the International Labor Organization (ILO), forced labor can be understood as "all work or service which is exacted from any person under the threat of a penalty and for which the person has not offered himself or herself voluntarily". Investigations have revealed how leading technology companies continue to benefit from forced labor. For instance, in 2021, seven of Apple's suppliers were found to be linked to suspected forced labor of Uyghur Muslims and other persecuted groups sourced from the Xinjiang region [7]. As suppliers of components often contract with multiple companies, it is estimated that the true extent of the ethical dilemma with the use of forced labor in the technology sector may be significantly underreported.

**Manufacturing sustainability:** Semiconductor companies use massive amounts of energy to manufacture chips. This energy consumption was shown to produce over 63% of the carbon emissions from manufacturing 12-inch wafers at TSMC. And the energy demand for next-generation manufacturing is expected to increase rapidly, e.g., up to 7.7 billion kilowatt-hours annually in a 3nm fab. Based on publicly available sustainability reports from AMD, Apple, Facebook, Google, Huawei, Intel, Microsoft, and TSMC, it has been reported that the hardware-manufacturing process, rather than system operation, is the primary source of carbon emissions [8]. In response, semiconductor companies are pledging to increasingly rely on renewable energy. For instance, TSMC is aiming to use renewable energy for 20% of its fabs' annual electricity consumption [9]. However, electronic manufacturing across multiple fabs across the globe is expected to continue to constitute a large portion of computing's global carbon footprint. Available data supports this trend. For example, the fraction of life-cycle carbon emissions due to hardware manufacturing increased from 49% for the iPhone 3GS to 86% for the iPhone 11 [8]. Beyond the indirect carbon footprint due to the use of "brown" energy for manufacturing, electronic chip manufacturing is also responsible for direct emissions from burning perfluorocarbons (PFCs), chemicals, and gases. TSMC indicates that nearly 30% of emissions from manufacturing 12-inch wafers are due to PFCs, chemicals, and gases [9]. The environmental costs of semiconductor manufacturing represent an important ethical consideration, one that must not be ignored as we progress further into the era of ubiquitous and connected computing.

## 2.2 IoT Design

The design of increasingly pervasive Internet of Things (IoT) computing devices involves many ethical concerns associated with the software design, hardware design, system integration, and post-deployment phase.

**Security and Privacy:** Protections against security and privacy attacks on computing systems require hardware and software modifications during the design process. The non-trivial overheads associated with integrating security and privacy protections (e.g., due to encryption/decryption protocols, key management systems, side-channel obfuscation techniques) can not only increase design costs and time-to-market, but also result in increased energy/power overheads and performance reduction. Thus, it is not surprising that in many low-cost computing chips and platforms, such as those found in various emerging

IoT applications, sufficient security and privacy countermeasures are absent, which raises serious ethical concerns. These IoT applications are particularly susceptible to being hacked and leak information. A 2019 study showed that over 90% of IoT communications across devices such as IP cameras, medical devices, industrial control systems, and 3D printers were unencrypted [10]. Smart TVs and smart home speakers with voice interaction capabilities frequently record conversations beyond commands intended for them [11]. Smart vehicles are increasingly using IoT-driven telemetry, infotainment, sensing, perception, and communication systems with vulnerabilities that have led to many well-publicized attacks, such as the 2014 Jeep Cherokee hack that was able to kill the vehicle engine while it was on a highway [12]. Toys with IoT devices such as Wi-Fi enabled Barbie dolls and robots have been hacked and turned into surveillance devices [13]. As IoT devices proliferate at near-exponential rates, such attacks will only become more widespread.

**Safety:** Many IoT systems are deployed in real-time and mission-critical contexts. If humans are involved in these use-cases. e.g., semi-autonomous self-driving vehicles and pacemakers, such systems must be designed with user safety as a primary design concern. However, inadequate design processes that do not anticipate corner cases in real-world deployments can often fail in ensuring user safety, raising serious ethical issues. Considering the medical application domain, the U.S. Food & Drug Administration (FDA) indicates that more than 80,000 deaths and 1.7 million injuries have been linked to faulty medical devices in the past decade. Trends from available data indicate that there has been a marked rise in medical device mishaps and recalls in recent years [14]. For example, in February 2016, 263,520 units of glucose monitoring (CGM) systems were recalled due to a faulty auditory alarm. The CGMs included a sensor placed subcutaneously to measure blood glucose readings in patients, which were then sent to the receiver. The faulty alarms remained inactive in the defective CGM systems during high or low blood glucose levels in patients, potentially leading to serious adverse events and even death. In June 2021, Philips recalled 3.5 million ventilator devices after finding a defect that could cause cancer. The ventilators used polyester-based polyurethane sound abatement foam, which had the potential to degrade into particles that could be ingested or inhaled and have toxic and carcinogenic effects. Inadequately factoring in safety considerations during system design can clearly have a significant impact, not just in medical contexts but also across many other safety-critical applications.

**Dealing with post-deployment issues:** Given the pressures of meeting stringent time-to-market goals, IoT platforms are often released without comprehensive validation at the software, hardware, and system levels. This can lead to many unintended bugs at these levels being discovered in the field. How companies respond to such situations is an important ethical concern. A classical example is Intel's Pentium bug in the 1990s. After the bug was observed during mathematical computations involving division operations in 1994, and disclosed in November of the same year, Intel admitted that that its own engineers had also discovered the Pentium's problems a few months earlier, but the company had decided that since encountering the error was so unlikely (it only affected decimal bits of lower significance during calculations), it would not need to notify Pentium customers [15]. After mass media picked up on the story, Intel decided upon a qualitative return policy. If a customer wanted a replacement chip, they would need to talk to people at Intel who would decide whether the customer really needed one. In December of the same year, IBM announced that it was halting shipments of its computers containing Pentium chips after they ran tests on their own and discovered that typical spreadsheet users might encounter a division error every 24 days, rather than every 27,000 years as Intel predicted. This prompted Intel to rescind its conditional replacement policy and offer replacement chips to anyone requesting one. Intel's CEO Andrew Grove stated: "Finally we decided, 'This is the right thing to do, both morally and ethically'". Intel's actions reflected their recognition that ethical practices and policies promote the long-term, best interests of a company. This cautionary tale serves to highlight how post-deployment issues must be carefully handled. Yet today, there are still too many instances of companies being aware of vulnerabilities in IoT devices and delaying informing consumers about them [16].

**Lifecycle carbon footprint:** IoT devices have a carbon footprint associated with their manufacture (as discussed earlier) and operation. Lifecycle assessment (LCA) of different battery-powered IoT devices (e.g., tablets, smartphones, wearables, laptops) and always connected devices (e.g., smart home speakers, desktops, game consoles) from Apple, Google, and Microsoft released after 2017 has revealed that manufacturing dominates emissions for battery-powered IoT devices, whereas operational energy consumption dominates emissions from always-connected devices [8]. Not surprisingly, the manufacturing footprint increases with increasing hardware capability (e.g., flops, memory bandwidth, storage). Unfortunately, the same study showed that software and hardware optimizations in recent years primarily focused on maximizing performance, while overlooking the trend with increasing carbon footprint [8]. With increasing IoT proliferation in wired and wireless contexts, the projected operational energy of IoT, which is currently about 5% of global energy demand, is expected to increase to 7% by 2030 [17]. The carbon emissions from the continued manufacture and operation of such devices therefore poses serious environment-related ethical challenges that must not be ignored.

**E-waste:** The end-of-life for IoT devices has ethical repercussions. E-waste, which refers to electronic products nearing the end of their useful life, has been doubling every few years and more than 90% of it being disposed illegally, according to the United Nations. A recent study indicated that more than 5 billion of the 16 billion mobile phones possessed worldwide will likely be discarded in 2022 [18]. If improperly disposed, e-waste can leach lead and other substances into soil and groundwater, which directly threatens human health and our environment. Discarded electronic devices are also openly burned in places like Agbogbloshie, Ghana and Guiyu, China to recover valuable metals such as copper, aluminum, and brass. The black and toxic fumes emitted from burning e-waste are harmful to anyone in the vicinity of such sites.

## 2.3 Artificial Intelligence (AI)

The increasing reliance on AI algorithms within IoT products (e.g., IP cameras with in-built object detection, wearable medical diagnosis devices) as well as computer design tools creates many ethical dilemmas.

**Transparency:** AI algorithms such as those based on deep neural networks represent black box approaches to solving problems, where both the learned mechanisms and the steps used to arrive at predictions cannot be easily explained, even by AI domain experts. This raises the question of how companies and third-party users of AI-driven systems can be transparent with customers that are inherently not entirely transparent. For example, in the medical IoT domain, if physicians cannot explain how an AI-based healthcare system arrived at a decision for a specific patient, to what extent should they rely on these solutions? Consider the use of Watson for Oncology which was widely used in China for health diagnosis via image recognition. It was later found that the underlying AI algorithms were primarily trained on a Western dataset leading to poor results for Chinese patients compared to Western patients [19]. Transparency (and "explainable AI") is thus a critical ethical requirement to prevent systemic misdiagnosis and other undesirable outcomes. But even developers may have a hard time explaining why their AI algorithms behave the way they do.

**Trust:** The increasing use of AI algorithms particularly in electronic design automation (EDA) tools used for chip design creates many ethical challenges. EDA tool vendors and chipmakers are increasingly using AI algorithms for design verification and simulation, logic synthesis, place-and-route, and timing and physical signoff analysis [20]. How can developers trust the outcomes of such AI tools, particularly given their lack of explainability? For example, can developers trust that AI algorithms for validation have covered typical and corner cases in designs effectively? There are also growing concerns related to backdoors and attacks (adversarial, poisoning) that can impact the quality of output, as well as security and privacy properties in chip design flows [21]. Such vulnerabilities can be introduced in design flows either by disgruntled or malicious employees or compromised supply chains with untrustworthy third-party algorithm developers. These developments make it difficult for both developers and consumers to trust that the designed chips will behave in a manner that promotes safety, privacy, and security.

**Bias:** If AI algorithms are trained on biased data, they may lead to undesirable outcomes that create ethical challenges. Biased AI algorithms can have serious implications, as highlighted by the case of an AI algorithm used by large healthcare systems and payers to guide health decisions for almost 200 million people in the U.S. annually. The algorithm incorrectly assigned the same level of risk to Black and White patients, despite Black patients in the dataset being much sicker [22]. The racial bias was a result of the algorithm using healthcare costs instead of illness as a measure of the level of health needs. As Black patients' healthcare related spending was lower, the algorithm incorrectly concluded that Black patients were healthier. Many other recent examples highlight similar biases, e.g., the Flickr mobile app's image recognition tool reportedly tagging black people as "animals" or "apes", Hewlett-Packard's software for web cameras struggling to recognize dark skin tones, and Nikon's camera software inaccurately identifying Asian people as blinking. The underlying AI-based algorithms in these systems learn by being fed certain images, often chosen by engineers, and the system builds a model of the world based on those images. If a system is trained on photos of people who are overwhelmingly white, it will have a harder time recognizing nonwhite faces. How developers can reduce bias in AI algorithms that are being increasingly integrated into smartphones, wearables, medical devices, robotics, automotive systems, and industrial automation, remains a pressing open problem.

## 3 ADDRESSING ETHICAL CHALLENGES

The landscape of ethical challenges facing semiconductor fabrication, IoT design, and AI integration is vast. Solutions to overcome the complex ethical problems outlined in the previous section will not be easy. Here we outline a few promising directions that can have a positive impact on enabling ethical design of computing systems.

**Transparency with sustainability data:** Clearly, reducing the carbon footprint and environmental impact of chipmaking and technology use is an important need. However, it is naive to expect a reduction in technology use or in the manufacture of IoT devices when all trends indicate an increase in technology proliferation in our everyday lives. The question then becomes: what can semiconductor manufacturers and IoT developers realistically do to reduce the carbon footprint of computing? Perhaps the most essential pre-requisite to even begin to address this problem is transparency from companies on the costs (direct and indirect) associated with their designs. A few large IT companies such as Apple, Amazon, Meta, Microsoft, and Google, have been publicly disclosing their carbon emissions, but more companies, including those involved in semiconductor manufacturing and IoT design need to follow suit. Moreover, greater transparency in reporting is also needed, to determine if salient factors have been accounted for in carbon footprint calculations. Lastly, sustainability reports need to go beyond energy data to capture impacts due to other resources involved, such as water usage and earth minerals.

**Sustainable design flows:** Industry can also make key changes to reduce the carbon footprint of hardware and software design. A greater reliance on renewables is perhaps the most direct approach to achieve this goal. Additionally, many other decisions during hardware and software design flows can make a positive impact. The use of sustainable resource managers that can migrate intensive workloads (e.g., logic synthesis runs, AI algorithm training) across geo-distributed data centers to better exploit availability of renewable energy, can not only be a strong incentive for companies (as it leads to cost savings) but also reduce carbon footprint of the migrated processes [23]. The use of custom and heterogeneous hardware to accelerate hardware and software design can reduce the energy consumption during development. For example, AI algorithm training on Google TPU processors could be much more energy-efficient than doing so on general purpose processors such as CPUs and GPUs. The use of such accelerators within IoT devices can also reduce the operational energy usage over the lifetime of the devices. Lastly, the design and manufacture of more reliable (hardened) components that have a longer endurance can extend the viable lifetime of IoT products and help limit e-waste.

**Programming ethical behaviors:** To ensure that IoT-based systems such as autonomous vehicles and medical devices behave ethically, it will be important to program ethical behaviors within the constituent hardware and software components. This can involve encoding the required ethical behavior explicitly in rules or creating algorithms to allow systems to determine appropriate ethical actions [24]. These rules can be based on ethical theories (e.g., deontology, teleology) and have an advantage in that they can be clearly understood by humans. Such rules can represent a range of the ethical behaviors that can be customized across application domains. As an example, [25] describes an approach for programming ethical behavior in autonomous vehicles by integrating ethical considerations into the costs and constraints used in automated control algorithms, to minimize damages in an incident. The rule-based approach can also be extended to allow for dynamic selection of rules based on context, and to provide device users the autonomy to make choices about ethical dilemmas, rather than have them be hard coded by developers.

**Maintaining security and privacy:** IoT platforms must be designed with clear policies to enable secure and privacy-preserving behavior. This can involve the use of data access control and sharing mechanisms, e.g., the design and integration of authentication and key establishment mechanisms, filters to mask shared personal data, user-configurable access rules for data handling, and mechanisms for digital anonymity. A combination of such approaches can help minimize data leakage and related risks. To mitigate security attacks, techniques for network access control (including firewalls), public or private key encryption, and authentication must be an integral part of IoT platforms. Justifying the cost of including these mechanisms in IoT devices requires a holistic cost-benefit that should include the costs related to damage in reputation, cost of recalls and replacements, and regulatory fines if vulnerabilities are detected and exploited during the product lifetime.

**Ethical AI:** If AI algorithms are integrated into IoT design flows or products, there is a need to enable transparency and mitigate bias. Ethical programming with such algorithms would then require that important decisions should be a 'white box' rather than a 'black box' so that stakeholders can scrutinize and understand how the algorithms make decisions and enable social accountability. Using open-source software can be one approach to achieve transparency and minimize bias. Getting algorithms or systems to explain their own actions and audit their own execution would be another approach (albeit very much a difficult and open problem today). The choice of dataset selection for training AI algorithms is also crucial to minimize bias. If there is significant class imbalance in the dataset, the AI algorithm can be easily biased towards the classes with greater representation in the dataset. Methods such as cost-sensitive class weighing, adaptive resampling, and re-collection of undersampled class data can be useful towards minimizing bias in such scenarios [26].

**Ethical LCA:** Ethical challenges can arise at many instances over the entire lifecycle of an IoT system. A careful ethics life cycle assessment (LCA) should include identification of, engagement with, and explicit communication about the diverse values and perspectives of all stakeholders – such as developers, sales reps, technicians for installation, users, deployment facilities (e.g., hospitals, vehicle manufacturing plants), and repair/debug specialists – while supporting systematic and thorough reflection and reasoning about the ethical issues. The reflection on ethical issues should go beyond impacts and consequences of using the IoT product and include considerations at all stages of the product lifecycle, including the earliest stages of initial conceptual design and market analysis (to determine the ethics of the multiple pathways to innovation), design, validation, deployment, lifecycle monitoring, repair, and retirement [24]. As an example, consider the ethical product lifecycle assessment from the agricultural biotechnology field [27]. The assessment advocates for the use of the Ethical Matrix method, which is a tool to evaluate the intersection of three normative ethical principles (respect for well-being, autonomy, and justice) with four relevant stakeholder groups (the treated organisms, producers, consumers, and environment). The applicability to IoT product development could be imagined with application of the same important principles to the relevant stakeholder across application domains. For example, for medical IoT devices, the stakeholders could include groups of patients, smart healthcare companies, surgeons, and hospitals, and concerns that encompass economic, regulatory, sustainability, and societal factors.

**Ethics-centric codes and regulations:** As discussed earlier, both the Institute of Electrical and Electronics Engineers (IEEE) and the Association of Computing Machinery (ACM), two of the computing field's largest professional associations, have published and revised codes of ethics. These codes are necessary to establish benchmarks for good practices and values, which is particularly important in helping those new to the profession to develop a moral professional compass. However, the code documents are brief and lack specific advice to address ethical dilemmas during the practical design and operational phases of IoT products. The reasons to comply with ethical codes are also often weak, and easily overridden by reasons to deviate from them, e.g., due to economic pressures. Therefore, regulatory support from the government (and in some cases, the specific profession) is crucial to incentivize making ethical decisions. Such regulatory support currently is emphasized only for safety-critical domains, e.g., healthcare. Considering the example of the healthcare domain, current regulatory frameworks used by the US FDA involve reviewing medical IoT devices through a premarket pathway, such as premarket clearance (510(k)), De Novo classification, or premarket approval. The FDA may also review and clear modifications to medical devices, including software as a medical device, depending on the significance or risk posed to patients of that modification. But emerging developments, such as with the increasing use of AI in medical contexts, creates challenges for regulatory frameworks that often cannot keep up with the rapid pace of change. The US FDA has acknowledged [28] that "The FDA's traditional paradigm of medical device regulation was not designed for adaptive artificial intelligence and machine learning technologies. Under the FDA's current approach to software modifications, the FDA anticipates that many of these artificial intelligence and machine learning-driven software changes to a device may need a premarket review." Thus, the challenge with regulations is that they are primarily developed and enacted as responses to already-existing ethical challenges and are often unable to address many of the ethical challenges that keep emerging as technology gets adopted and used in new ways. Nonetheless, despite such shortcomings, regulations remain the only viable mechanism to hold erring businesses accountable. Regulations can also be combined with public policy to target bigger issues, e.g., in the healthcare domain, regulations and public policy can aim for fair distribution of healthcare benefits and protecting equality of care in society. Such regulation and public policy needs to expand beyond safety-critical application domains to encompass the ethical challenges emerging in non-safety critical applications, e.g., data privacy violations in wearable and mobile IoT devices.

**Ethics-centric workforce education:** In response to the bad publicity generated from IoT device mishaps and recalls, e.g., in the autonomous vehicle and medical device application domains, efforts are being made to improve general awareness of ethical concerns for engineers and scientists involved in research and design of IoT technologies. Many universities and research institutions are beginning to emphasize topics related to ethics in their technical curricula with the goal of raising ethical awareness in developers, programmers, and engineers. Many technology companies working with AI are implementing training modules on ethics for their employees, e.g., ethical foresight analysis, to educate designers and managers in predicting potential ethical issues and the consequences of specific technologies. These are steps in the right direction. But many open problems remain. Integrating ethical topics in curricula that are already packed with courses and with little wiggle room remains a challenge. It is also not always clear how to go about identifying ethical dilemmas associated with emerging technologies, especially if their usage modalities are unconventional or without precedent. At the very least, in both educational and industry ethics training modules, emphasis should be placed on practical case studies to highlight the appropriateness of mapping various ethical theories (e.g., deontological, utilitarian) to a particular situation, as part of applied ethics analyses.

## 4 ETHICS-AWARE DESIGN METHODOLOGIES

The development of methodologies to guide analysis and decision-making at different levels of abstraction is a natural one for computer scientists and engineers. Ethical design of computer systems will require integrating the approaches described in the previous section into existing design methodologies. Ethics can be incorporated into design methodologies at several levels of abstraction:

- System architectures can be evaluated for their effectiveness in achieving ethical goals: responsible materials, manufacturing, reliability, bias, *etc*. McFarland [29] suggested roles for non-technical participants as members of ethical reviews, particularly in the early stages of a project.
- Components can be designed to standards of reliability, safety, and lack of bias. Design reviews and code inspections are commonly used in software design [30]. Ethical considerations can be addressed as one component of an overall design review. Design reviews typically incorporate experts from multiple aspects of the design; ethics experts can be included on the team to ask ethics questions and evaluate the completeness and appropriateness of the design's response.
- Testing can measure the effectiveness of ethical methods at all levels of abstraction from unit testing to system testing. For example, bias can be measured as part of the testing process.
- Materials, components, and manufacturing processes can be specified to take into account ethical requirements on materials, processes, and worker safety.

Thus, traditional design methodologies must change to include ethics-centric goals, in additional to traditional design goals such as performance, energy-efficiency, cost, form factors, etc. This is because the costs of ignoring ethics in computer design are becoming greater than ever before, with the increasing reliance on computers in every facet of our lives, and particularly in safety-critical systems. The ethics-centric methodologies must be applied at many different points in the lifetime of a system or the career of a computing professional: initial design, detailed design, implementation and construction, deployment, maintenance, decommissioning. Computing professionals must recognize the importance of their ethical obligations in these methodologies, which should range from identifying and preventing problems to reporting and whistleblowing if ethical concerns are uncovered.

## 5 CONCLUSIONS

In this article, we provided a broad perspective on the ethical challenges within semiconductor chip design, IoT applications, and the increasing use of AI in the design processes, tools, and hardware-software stacks of these systems. We discussed important ethical challenges associated with the use of conflict minerals, semiconductor fabrication toxins, forced labor, environmentally sustainable manufacturing, security/privacy, safety, post-deployment issues, IoT lifecycle carbon footprint, e-waste, and emerging challenges related to transparency, bias, and trust with the use of AI in IoT systems. All of these ethical challenges are deeply intertwined during the hardware and software design, and operation of IoT platforms, whose use is growing at an exponential rate. We advocate for addressing these ethical challenges on multiple fronts: greater data transparency and sustainable design efforts from companies, programming of ethical behaviors, integration of more effective security/privacy mechanisms, ethical AI algorithm design, improved ethics-centric lifecycle assessment, combining regulations with public policy, and better workforce education. Such efforts are crucial to establish a multi-pronged framework for realizing ethical manufacturing and operation of computing systems.

## ACKNOWLEDGMENTS

Pasricha's work was supported in part by grant CNS-2132385 from the National Science Foundation (NSF). Wolf's work was supported in part by NSF grant 2002854.

**Sudeep Pasricha** (sudeep@colostate.edu) received his Ph.D. in computer science from the University of California, Irvine in 2008. He is currently a Professor at Colorado State University. His research interests include networks-on-chip, and hardware/software co-design for energy-efficient, secure, and fault-tolerant embedded systems. He is a Senior Member of IEEE.

**Marilyn Wolf** (mwolf@unl.edu) is Koch Professor of Engineering and Director of the School of Computing at the University of Nebraska – Lincoln. Her research interests include embedded computing, cyber-physical systems, Internet-of-Things systems, and embedded computer vision. She is a Fellow of the IEEE.

Mail Address : 1373 Campus Delivery, Colorado State University, Fort Collins, CO 80523-1373